\documentclass[12pt]{iopart}
\usepackage{iopams}
\expandafter\let\csname equation*\endcsname\relax
\expandafter\let\csname endequation*\endcsname\relax
\usepackage{amsmath}
\pdfoutput=1
\usepackage{graphicx,psfrag,color}
\usepackage[utf8]{inputenc}
\usepackage[T1]{fontenc}
\usepackage{lmodern}
\usepackage{times}
\usepackage{mathptmx}
\usepackage{epstopdf}
\usepackage{xcolor,physics}
\usepackage[colorlinks,linkcolor=blue,urlcolor=blue,citecolor=blue]{hyperref}
\usepackage[normalem]{ulem}
\usepackage{xspace}
\usepackage{float}
\usepackage{bm}
\usepackage{cite}
\usepackage{lineno}
\hypersetup{
    colorlinks=true,
    linkcolor=blue,
    filecolor=magenta,
    urlcolor=cyan,
    citecolor=cyan
}

\def\la{\langle}
\def\ra{\rangle}

\DeclareSymbolFont{bbold}{U}{bbold}{m}{n}
\DeclareSymbolFontAlphabet{\mathbbold}{bbold}
\makeatletter
\newcommand*\bigcdot{\mathpalette\bigcdot@{.55}}
\newcommand*\bigcdot@[2]{\mathbin{\vcenter{\hbox{\scalebox{#2}{$\m@th#1\bullet$}}}}}
\makeatother

\def\im{\mathrm{i}}
\def\ee{\mathrm{e}}
\def\bra#1{\mathinner{\langle{#1}|}}
\def\ket#1{\mathinner{|{#1}\rangle}}
\providecommand{\keywords}[1]{\textbf{\textit{Keywords:}} #1}

\begin{document}

\newcommand{\gae}{\lower 2pt \hbox{$\, \buildrel {\scriptstyle >}\over {\scriptstyle
\sim}\,$}}
\newcommand{\lae}{\lower 2pt \hbox{$\, \buildrel {\scriptstyle <}\over {\scriptstyle
\sim}\,$}}

\title{Stochastic resets in the context of a tight-binding chain driven by an oscillating field}
\author{Sushanta Dattagupta$^1$,  Debraj Das$^2$, and Shamik Gupta$^3$}
\address{$^1$National Institute of Technology, Mahatma Gandhi Road, Durgapur, West Bengal 713209, India \\ $^2$Department of Engineering Mathematics, University of
Bristol, Bristol BS8 1TW, United Kingdom\\  
$^3$Department of Theoretical Physics,  Tata Institute of Fundamental Research,  Homi Bhabha Road,  Mumbai 400005, India}
\ead{sushantad@gmail.com,debrajdasphys@gmail.com,shamikg1@gmail.com}
\date{\today}
\begin{abstract}  
In this work, we study  in the framework of the so-called driven tight-binding chain (TBC) the issue of quantum unitary dynamics interspersed at random times with stochastic resets mimicking non-unitary evolution due to interactions with the external environment, The driven TBC involves a quantum particle hopping between the nearest-neighbour sites of a one-dimensional lattice and subject to an external forcing field that is periodic in time. We consider the resets to be taking place at exponentially-distributed random times. Using the method of stochastic Liouville equation, we derive exact results for the probability at a given time for the particle to be found on different sites and averaged with respect to different realizations of the dynamics.  We establish the remarkable effect of localization of the TBC particle on the sites of the underlying lattice at long times.  The system in the absence of stochastic resets exhibits delocalization of the particle, whereby the particle does not have a time-independent probability distribution of being found on different sites even at long times, and, consequently,  the mean-squared displacement of the particle about its initial location has an unbounded growth in time.  One may induce localization in the bare model only through tuning the ratio of the strength to the frequency of the field to have a special value,  namely, equal to one of the zeros of the zeroth order Bessel function of the first kind.  We show here that localization may be induced by a far simpler procedure of subjecting the system to stochastic resets. 
\end{abstract}
\date{\today}
\maketitle
\tableofcontents

\section{Introduction}
\label{sec:intro}

In recent years, there has been an upsurge of interest in the study of quantal evolution in the presence of a dissipative environment. In condensed matter physics, the environment is often in the form of a heat bath comprising bosonic or fermionic excitations, giving rise to a subject called dissipative quantum mechanics~\cite{Weiss,Dattagupta}. Yet another distinct direction of focus is the case of an environment in the guise of a classical apparatus. The apparatus, when brought in contact with the system of interest, inevitably causes disruption in the mould of non-unitary evolution of an otherwise unitary dynamics. The case in point is that of quantum measurements or that of stochastic resets of quantum dynamics~\cite{Mukherjee}. While a lot of attention has been devoted to quantum measurement issues, our focus of attention here would be in the latter.   Stochastic resets in classical and quantum systems have been a subject of intense research in recent years, resulting in a number of interesting results of both theoretical and practical relevance.  For an extensive review on the subject, the readers are referred to Ref.~\cite{Satya-review}.
        
        At this stage, it is pertinent to point out that the effect of environment assumes heightened prominence when the size of the system of interest is ‘small’. The relevant instances are those of nano mesoscopic systems in which the surface-to-volume ratio is substantially ‘large’. Thus, while such systems are of great interest in their ability to preserve quantum coherence and hence, store quantum information, the flip side is the inevitable disruptive influence of the environment. In this paper, we consider such a situation in which the environment triggers stochastic sequences of resetting, at random instants of time, when the underlying density operator of the system of interest is reset to its form just prior to switching-on of the environmental coupling. This ongoing renewal process continues over a long time, and our interest is to see what it does to the form of the density operator from which physical quantities can be calculated. Needless to say, in-between the aforesaid resetting events, the system of interest is endowed with unitary evolution. However, resetting renders the underlying evolution non-unitary. 
          
        Because the effects under consideration are most prominent for small systems, we employ the much-studied tight-binding chain (TBC) of solid state physics~\cite{Dunlap:1986}. Such a one-dimensional toy model aptly describes quantum transport phenomena in a nano wire~\cite{Dattagupta-resonance}. The TBC, in the absence of any external force-field, is the discrete version of the free space dynamics of a quantum particle, e.g., an electron, which can be exactly handled in terms of what are called Bloch states that play the role of plane wave states of a free particle. In the TBC model, the particle, though localized most of the times on lattice sites, can occasionally tunnel to a nearest neighbour site with a frequency $\Delta$.     
        
        In recent times, we ourselves have considered the paradigmatic case of free tunnelling motion in the TBC in the context of projective quantum measurements as well as stochastic resets in Ref.~\cite{Das:2022}, hereafter designated as I.  In I, we encountered an interesting effect of localization of the TBC particle on the sites of the underlying lattice, whereby the free particle-like unbounded motion gets localized because of random interruptions occasioned by the resetting process.  This phenomenon of localization will occupy a significant space in our subsequent discussions. 
        
        It is important now to underscore that the TBC we treat here is not a free one as investigated in I, but is subject to an external perturbation in the form of an oscillatory force field. It is therefore appropriate at this stage to emphasize the new physics that ensues when the external perturbation is a time-periodic one. But, before that, assume the force to be static, with magnitude given by $F_0$ on every site of the TB lattice. While a particle in open space would feel a bias in motion, the space-periodic nature of the TB lattice gives rise to an interesting phenomenon called Bloch oscillations. This implies that the particle, starting from a given site of the lattice, keeps revisiting that site periodically at a frequency determined by $F_0$. Consequently, the effect is also dubbed as Wannier-Stark localization~\cite{Wannier,Dattagupta-resonance}. The effect however dramatically changes when the external force is an oscillatory one: $F_0 \cos(\omega t)$~\cite{Dunlap:1986}. Now, whenever the time $t$ matches with $2\pi n/\omega$,  $n$ being an integer, the Wannier-Stark localization gives way to delocalization, as though the particle is like a free particle, albeit the underlying tunnelling is governed by an effective frequency $\Delta_\mathrm{eff}$. A further fascinating phenomenon ensues at certain ratios of $F_0/\omega$ (namely,  when the ratio equals one of the zeros of the zeroth order Bessel function of the first kind), when $\Delta_\mathrm{eff}$ vanishes, yielding dynamic localization.
        
        Given this background, our aim in this paper would be to carefully delineate the concomitant presence of delocalization/localization and localization induced by stochastic resets, and make a detailed analysis of their interplay. Needless it is to point out that dynamic localization is a coherent phenomenon governed by the quantum phase, while localization due to stochastic resets carries the signature of decoherence due to non-unitary interventions of the surroundings. This coherence-to-decoherence cross-over or transition, not hitherto studied in the context of stochastic resets in quantum systems, is a recurring theme in the contemporary investigations of quantum information processes. 
        
The central result of out work is contained in Eq.~(\ref{eq:result-rho}), which gives the averaged probability for the TBC particle to be on site $m$ at time $t$, while  starting at site $n_0$ and evolving under the TBC Hamiltonian (\ref{eq:TBC-Hamiltonian-field}) with repeated stochastic resets at exponentially-distributed random times. From Eq.~(\ref{eq:result-rho}) follows the mean-squared displacement (MSD) about the initial location, given by $\overline{S}(t) = \langle (m-n_0)^2\rangle$, which as $t \to \infty$ at a fixed $\lambda$ approaches a time-independent non-zero value given by Eq.~(\ref{eq:MSD-asymptotic}).
 The latter result suggests localization of the particle on the sites of the lattice at long times. This may be contrasted with the behaviour in the absence of resets, when the particle has an unbounded growth of the MSD in time, with no signatures of localization. Figure~\ref{fig:Sbar-vs-t-Bessel0} depicts a coherence-to-decoherence cross-over, namely, dynamic localization of the bare model (i.e., the TBC subject to a time-dependent field, but with no resets ($\lambda=0$)) crossing over to localization in presence of stochastic resets when resets are sufficiently frequent (large $\lambda$); correspondingly, the MSD crosses over from a time-dependent to a time-independent value. This phenomenon of localization through stochastic resets when the TBC is subject to a time-dependent external field constitutes a major physical implication of our analysis presented in this work. 
        
        With these preliminary remarks on the scope and objectives of this paper, the rest of the manuscript is structured as follows. In Section~\ref{sec:model-no-reset}, we present an overview of the known results for a field-driven TBC in the absence of resets,  derived using the interaction picture of time evolution in quantum mechanics, differently from what is known in the literature~\cite{Dunlap:1986}.  Section~\ref{sec:model-reset} constitutes the core of the paper containing the main results for the field-driven TBC in the presence of stochastic resets.  Our numerical results illustrating the combined occurrence of delocalization/localization of the bare model and localization induced by stochastic resets are included in Section~\ref{sec:numerics}. Finally, Section~\ref{sec:conclusions} concludes the paper with our principal conclusions. In the two Appendices, we recall the notion of a superoperator used extensively in the main text and the issue of time evolution of the density operator under a time-dependent Hamiltonian.

\section{The field-driven TBC in the absence of resets}
\label{sec:model-no-reset}

In this section, we gather known results on the behaviour of the tight-binding chain (TBC) when driven by an external forcing field that is periodic in time~\cite{Dunlap:1986}. The results presented in this section will be used in the later part of the paper when we consider the effects of stochastic resets in the set-up.  In contrast to the analysis of Ref.~\cite{Dunlap:1986},  we find it convenient to obtain the results presented therein by using the interaction picture of time evolution in quantum mechanics~\cite{Messiah}.  Let us now proceed to discuss the analysis and the results.

The Hamiltonian of the TBC that describes a quantum particle hopping between the nearest-neighbour sites of a one-dimensional lattice is given by~\cite{Dunlap:1986}
\begin{align}
H_\mathrm{TBL}  = - \frac{\hbar \Delta}{2} \sum_{n=-\infty}^\infty \Big( |n\rangle  \langle n+1| + |n+1\rangle \langle n| \Big),
\label{eq:TBC-Hamiltonian}
\end{align}
where $\Delta > 0$ is the nearest-neighbour intersite hopping integral (the so-called tunnelling frequency),  and $|n\rangle$'s denote the so-called Wannier states that represent the state of the particle when located on site $n$.  These states form a complete basis satisfying $\langle m|n\rangle=\delta_{m,n}$ and $\sum_{m=-\infty}^\infty |m\rangle \langle m|=\mathbb{I}$.  In this work,  we will take the lattice spacing to be unity without loss of generality,  and set the Planck's constant to unity.  In terms of the operators
\begin{align}
K \equiv \sum_{n=-\infty}^\infty |n\rangle  \langle n+1|, ~~K^\dagger \equiv \sum_{n=-\infty}^{\infty} |n+1\rangle \langle n|,   
\end{align}
we have      
\begin{align}   
H_\mathrm{TBL} =-\frac{\Delta}{2}  (K + K^\dagger). 
\label{eq:model-1}
\end{align}                                                                  
The operators $K$ and $K^\dagger$ satisfy $KK^\dagger=K^\dagger K=\mathbb{I}$ with $\mathbb{I}$ being the identity operator,  so that $[K,K^\dagger]=0$.  The latter fact implies that there exists a representation in which $K$ and $K^\dagger$ can be simultaneously diagonalized. Such a representation is given by the so-called Bloch states, which are defined in the reciprocal momentum-space by
\begin{align}
|k\rangle \equiv \frac{1}{\sqrt{2\pi}}\sum_{n=-\infty}^\infty \ee^{-\mathrm{i}nk}|n\rangle.
\end{align}
Note that $|n\rangle = (1/\sqrt{2\pi})\int_{-\pi}^\pi \mathrm{d}k\, \exp(\mathrm{i}nk)|k\rangle$.  One has 
\begin{align}
K |k\rangle = \ee^{-\mathrm{i}k} |k\rangle, ~~K^\dagger |k\rangle = \ee^{\mathrm{i}k} |k\rangle,
\label{eq:Bloch1}
\end{align}             
so that the Bloch state $|k\rangle$ is an eigenstate of $K$ and $K^\dagger$, with eigenvalues $\exp (- \mathrm{i} k)$ and $\exp (\mathrm{i} k)$, respectively.  Moreover, one has $\langle k'|k\rangle=\delta(k-k')$.

We want to study the case when the system~(\ref{eq:model-1}) is subject to an external forcing field $F(t)$ that is periodic in time:  $F(t)=\mathcal{F}_0\cos (\omega t)$, with $T=2\pi/\omega$ being the time period of variation of the field.  In the context of a.c. electrical conductivity measurement due to an applied electric field with amplitude $E$, one has $\mathcal{F}_0=qE$, with $q=-e$ being the charge of the quantum particle contributing to electrical conductivity, i.e.,  electron.  The TBC Hamiltonian in presence of the field is explicitly time-dependent and has the form
\begin{align}
H=H(t) = H_\mathrm{TBL}+H_0(t);~~H_0(t)=F_0 \cos(\omega t)N,
\label{eq:TBC-Hamiltonian-field}
\end{align}
in terms of an operator $N$ given by
\begin{align}
N \equiv \sum_{n=-\infty}^\infty n |n\rangle \langle n|;~~N^\dagger=N.
\end{align}
The operator $N$ being diagonal in the Wannier states and not in the Bloch states causes transitions among the latter states. It is easily checked that one has
\begin{align}
[K,N]=K,~~[K^\dagger,N]=-K^\dagger,
\label{eq:KKdagger-commutation}
\end{align}
and so the two terms in the Hamiltonian~(\ref{eq:TBC-Hamiltonian-field}) do not commute with each other. We will refer to $F_0$ as the strength of the forcing field. Note that $H(t)$ for two different times do not commute: $[H(t),H(t')]=(\Delta/2)F_0(\cos(\omega t)-\cos(\omega t'))(K-K^\dagger)$, while we have $[H_0(t),H_0(t')]=0$.
        
Let us investigate the question: what is the probability $P_m(t)$ that the particle is on site $m$ at time $t>0$, given that the particle was on site $n_0$ at time $t=0$? If $\rho(t)$ denotes the density operator of the system at time $t$, we have by definition that
\begin{align}
P_m(t)=\langle m|\rho(t)|m\rangle. 
\label{eq:Pmt-as-rho-element}
\end{align}
The time evolution of $\rho(t)$ is given by the quantum Liouville equation 
\begin{align}
\frac{\partial \rho(t)}{\partial t}=-\im \big[ H(t), \rho(t)\big].
\label{eq:liouville-eqn}
\end{align} 
Equation~\eqref{eq:liouville-eqn} has
the formal solution (see Appendix~A):
\begin{align}
\rho(t)=\ee_+^{-\im \int_0^t \mathrm{d}t'~H(t')}\,\rho(0) \, \ee_-^{\im \int_0^t \mathrm{d}t'~H(t')},
\label{eq:rho-evolution}
\end{align}
where because $H(t)$ at two different times do not commute with each other that we invoked time ordering in writing down the exponential factors in the above equation; the minus and plus subscripts on the exponential indicate negative (i.e., the latest time to the right) and positive (i.e., the latest time to the left) time ordering, respectively.  Equation~(\ref{eq:rho-evolution}) implies unitary evolution of the density operator that preserves its trace: $\mathrm{Tr}[\rho(t)]=1~\forall~t$. 

In order to obtain $P_m(t)$, it is instructive to invoke the interaction picture of time evolution in quantum mechanics and transform $\rho(t)$ to $\widetilde{\rho}(t)$ given by
\begin{align}
\widetilde{\rho}(t)\equiv \ee^{\im \int_0^t \mathrm{d}t'~H_0(t')}\rho(t) \, \ee^{-\im \int_0^t \mathrm{d}t'~H_0(t')}, 
\label{eq:rho-transformation}
\end{align}
 where note that $H_0(t)$ at two different times commute with each other so that no time ordering is required in writing down the exponential factors.  Since $H_0(t)$ is diagonal in the Wannier states,  we have $P_m(t)=
\langle m|\rho(t)|m\rangle=\langle m|\widetilde{\rho}(t)|m\rangle$. 

 Using the Liouville equation~(\ref{eq:liouville-eqn}), we get
\begin{align}
\frac{\partial \widetilde{\rho}(t)}{\partial t}=-\im[H_\mathrm{TBL}(t),\widetilde{\rho}(t)],
\label{eq:rhotilde}
\end{align}
with
\begin{align}
H_\mathrm{TBL}(t) \equiv \ee^{\im \int_0^t \mathrm{d}t'~H_0(t')}H_\mathrm{TBL} \, \ee^{-\im \int_0^t \mathrm{d}t'~H_0(t')}.
\label{eq:HTBLt-0}
\end{align}

Now,  using the Baker-Campbell-Hausdorff formula given by $\ee^A B\, \ee^{-A}=B+[A,B]+(1/2!)[A,[A,B]]+(1/3!)[A,[A,[A,B]]]+\ldots$,  we may obtain from Eq.~(\ref{eq:HTBLt-0}) on using Eq.~(\ref{eq:KKdagger-commutation}) that
\begin{align}
H_\mathrm{TBL}(t)&=V^+ \cos\left((F_0/\omega)\sin (\omega t)\right)-\im V^-\sin\left( (F_0/\omega)\sin (\omega t)\right) \nonumber \\
&=-\frac{\Delta}{2}\left(K \ee^{-\im \eta(t)}+K^\dagger \ee^{\im \eta(t)}\right),
\label{eq:HTBLt}
\end{align}
with $\eta(t) \equiv (F_0/\omega) \sin (\omega t)$, 
and we have introduced the operators
$V^\pm\equiv -(\Delta/2)(K \pm K^\dagger)$.  Note that $H_\mathrm{TBL}(t=0)=(-\Delta/2)(K+K^\dagger)=H_\mathrm{TBL}$, as expected. It then follows that $[H_\mathrm{TBL}(t),H_\mathrm{TBL}(t')]=0$.  

Equation~(\ref{eq:rhotilde}) has the solution
\begin{align}
\widetilde{\rho}(t)=\ee^{-\im \int_0^t \mathrm{d}t'~H_\mathrm{TBL}(t')}\rho(0) \, \ee^{\im \int_0^t \mathrm{d}t'~H_\mathrm{TBL}(t')},
\label{eq:rhotilde-solution}
\end{align}
where we have used the fact that $H_\mathrm{TBL}(t)$ at two different times commute with each other.  We therefore have
\begin{align}
P_m(t)=\langle m|\ee^{-\im \int_0^t \mathrm{d}t'~H_\mathrm{TBL}(t')}|n_0\rangle \langle n_0|\ee^{\im \int_0^t \mathrm{d}t'~H_\mathrm{TBL}(t')}|m\rangle,
\label{eq:Pmt-definition}
\end{align}
where we have used the fact that $\rho(0)=|n_0\rangle \langle n_0|$.
Comparing Eq.~(\ref{eq:rhotilde-solution}) with Eq.~(\ref{eq:rho-transformation}) and using Eq.~(\ref{eq:rho-evolution}),  we get
\begin{align}
\ee^{-\im \int_0^t \mathrm{d}t'~H_\mathrm{TBL}(t')}=\ee^{\im \int_0^t \mathrm{d}t'~H_0(t')}\ee_+^{-\im \int_0^t \mathrm{d}t'~H(t')}.
\label{eq:HTBLt'-relation}
\end{align}

Equation~(\ref{eq:HTBLt}) yields 
\begin{align}
\ee^{\im \int_0^t \mathrm{d}t'~H_\mathrm{TBL}(t')}= \ee^{-\im (\Delta/2)[Kw^\star(t)+K^\dagger w(t)]},
\label{eq:expHTBLt}
\end{align}
with
\begin{align}
&w(t) \equiv \int_0^t \mathrm{d}t'~ \ee^{\im \eta(t')}=\frac{1}{\omega}\int_0^{\omega t}\mathrm{d}\tau~ \ee^{\im (F_0/\omega)\sin \tau}.
\end{align}
Using the substitution $t=\ee^{\im \tau}$ in the definition of the generating function of the Bessel function of the first kind~\cite{DLMF-NIST}: $
\ee^{(x/2)(t-1/t)}=\sum_{p=-\infty}^\infty J_p(x)t^p$,  one has $\ee^{\im x \sin \tau}=\sum_{p=-\infty}^\infty J_p(x) \, \ee^{\im p\tau}$, so that 
\begin{align}
w(t)&=\frac{1}{\omega}\int_0^{\omega t}\mathrm{d}\tau~\sum_{p=-\infty}^\infty J_p(F_0/\omega) \ee^{\im p \tau}.
\label{eq:wt-Bessel}
\end{align}
When $t=nT$,  i.e.,  $\omega t=2\pi n$,  with $n$ being an integer,  the integral over $\tau$ vanishes unless $p=0$, and one gets
\begin{align}
w(nT)=tJ_0(F_0/\omega).
\label{eq:wmT}
\end{align}

From Eq.~(\ref{eq:Pmt-definition}), we get on transforming to the Bloch states and using Eqs.~(\ref{eq:expHTBLt}),  (\ref{eq:Bloch1}) and the result $\langle k'|k\rangle=\delta(k-k')$ that
\begin{align}
P_m(t) &=\frac{1}{(2\pi)^2}\int_{-\pi}^\pi \mathrm{d}k\int_{-\pi}^\pi \mathrm{d}k'~ \ee^{\im (m-n_0)(k'-k)} \ee^{\im \Delta[(\cos k-\cos k')u(t)-(\sin k-\sin k')v(t)]} \nonumber \\
&=\Big|\frac{1}{2\pi}\int_{-\pi}^\pi \mathrm{d}k~\ee^{-\im (m-n_0)k+\im \Delta(u(t)\cos k-v(t)\sin k)}\Big|^2.
\label{eq:Pmt-0}
\end{align}
Here, we have defined the real quantities $u(t) \equiv \mathrm{Re}(w(t))$ and $v(t) \equiv \mathrm{Im}(w(t))$.
Using the identities~\cite{Dunlap:1986} $\ee^{-\im Vu(t)\cos k}=\sum_{n=-\infty}^\infty \ee^{-\im n\pi/2} \, \ee^{\im nk}J_n(Vu(t))$, $\ee^{-\im Vu(t)\sin k}=\sum_{n=-\infty}^\infty \ee^{\im nk}J_n(Vu(t))$ lets Eq.~(\ref{eq:Pmt-0}) be written straightforwardly on using $\int_{-\pi}^\pi \mathrm{d}k~\ee^{\im nk}=2\pi \, \delta_{n,0}$ as
\begin{align}
P_m(t)=\Big|\sum_{n_1=-\infty}^\infty \ee^{\im n_1 \pi/2} J_{n_1}(\Delta u(t))J_{m-n_0-n_1}(\Delta v(t))\Big|^2.
\end{align}
We next use the Graf's addition theorem for Bessel functions~\cite{Dunlap:1986}:
\begin{align}
\sum_{n=-\infty}^\infty \ee^{-\im n\pi/2} J_n(u(t))J_{m-n}(v(t))&=(-\Lambda)^m J_{-m}\bigg(\sqrt{u^2(t)+v^2(t)}\bigg) =\Lambda^m J_m\bigg(\sqrt{u^2(t)+v^2(t)}\bigg),
\end{align}
with $\Lambda \equiv \sqrt{(v(t)-\im u(t))/(v(t)+\im u(t))}$; here, we have used $J_m(x)=(-1)^m J_{-m}(x)$.  Note that we have $|\Lambda|^2=1$.  We therefore have 
\begin{align}
P_m(t)=\Big|(\Lambda^\star)^{m-n_0} J_{m-n_0}(\Delta |w(t)|)\Big|^2=J^2_{m-n_0}(\Delta |w(t)|).
\label{eq:Pmt-final}
\end{align}
It is easily checked that in the absence of forcing, when one has $w(t)=t$, the above equation reduces to the following result known in the literature (see, e.g., ~\cite{Das:2022}): $P_m(t)=J_{m-n_0}^2(\Delta t)$.

Note that since one has $J_{-m}(x)=(-1)^mJ_m(x)$,  the quantity $P_m(t)$ in Eq.~(\ref{eq:Pmt-final}) is symmetric under $m-n_0 \to -(m-n_0)$. It then follows that the average displacement of the particle from its initial location is
\begin{align}
\mu(t) \equiv \langle (m-n_0)\rangle = \sum_{m=-\infty}^\infty (m-n_0)P_m(t)=0,
\label{eq:MEAN-lam0-field}
\end{align}
so that $\langle m \rangle=n_0$.  In other words, the average position of the particle at any instant is its initial location.  Next, using the result~\cite{Dunlap:1986} $\sum_{m=-\infty}^\infty m^2J_m(x)=x^2/2$, we get the mean-squared displacement (MSD) as 
\begin{align}
S(t) \equiv \langle (m-n_0)^2 \rangle = \sum_{m=-\infty}^\infty (m-n_0)^2 P_m(t)=\frac{\Delta^2 |w(t)|^2}{2}.
\label{eq:MSD-lam0-field}
\end{align}
For the particular case of no field,  implying $w(t)=t$, we recover using the above equation the known result~\cite{Dunlap:1986,Das:2022}: $S(t)=\Delta^2t^2/2$.  From Eq.~(\ref{eq:MSD-lam0-field}), we get on using Eq.~(\ref{eq:wmT}) that
\begin{align}
S(t=nT)=\frac{\Delta^2 t^2 J_0^2(F_0/\omega)}{2}.
\label{eq:MSD-lam0-field-specialt}
\end{align}
We thus see that whenever the time of observation $t$ equals an integer multiple of the time period $T$ of the external periodic forcing,  the MSD has the free-particle-like ballistic behaviour seen in the unforced system,  $S(t) \propto t^2$, a difference being that one has in the forced system an effective tunnelling frequency 
\begin{align}
\Delta_\mathrm{eff}\equiv \Delta J_0(F_0/\omega)
\end{align}
 that is smaller than the parent tunnelling frequency $\Delta$ (since $|J_0(x)|\le 1$). The phenomenon of an MSD growing unbounded in time, as seen in Eqs.~(\ref{eq:MSD-lam0-field}) and (\ref{eq:MSD-lam0-field-specialt}),  implies delocalization, whereby the particle delocalizes about its initial location through the application of the external periodic field~\cite{Dunlap:1986}.  However,  if we choose $F_0/\omega$ to be such that the ratio equals a zero of $J_0(x)$ (i.e.,  $F_0/\omega$ is such that $J_0(F_0/\omega)=0$),  then we have $S(t=nT)=0$, and, moreover, $S
 (t)$ is bounded in time~\cite{Dunlap:1986}.  In other words,  in such a situation,  the MSD of the particle about its initial location is zero whenever $t=nT$, a phenomenon that is referred to as dynamic localization~\cite{Dunlap:1986}.
   
\section{The field-driven TBC in presence of resets}
\label{sec:model-reset}
 
In this section, we briefly discuss following Ref.~\cite{Das:2022} our formalism of Stochastic Liouville equation to study analytically unitary evolution interspersed at random times with instantaneous non-unitary interactions in the context of a general quantum system, and then apply the formalism to obtain explicit results for the TBC subject to a periodic forcing field in the case in which the interactions implement stochastic resets.  The quantum system while starting with a density operator $\rho(0)$ is assumed to be undergoing unitary evolution in
time for a fixed time $t$. The evolution is interspersed with instantaneous interaction with the environment at random times, which induces non-unitarity in the evolution of the system, and which is implemented by a given interaction operator $\mathcal{T}$. Thus,  a realization of the  evolution over the time interval $[\,0,\,t\,]$ may involve $\mu \ge 0$ number of instantaneous interactions occurring at
random time instances $t_1,t_2,\ldots,t_\mu$, where the times $\tau_{\nu+1} \equiv t_{\nu+1} -t_{\nu};~\nu=0,1,2,\ldots,\mu-1; \,t_0=0$ between
successive interactions are random variables that we choose to  be sampled independently from an exponential distribution:
\begin{align}
p(\tau)=\lambda\ee^{-\lambda\tau}.
\label{eq:ptau-exponential}
\end{align}
Here,  $\lambda>0$ is the inverse of the average time between two
successive interactions.  The evolution ends with unitary evolution for time duration $t - t_\mu$. We will later in the paper consider the form of $\mathcal{T}$ that implements stochastic resets of quantum dynamics~\cite{Mukherjee,Das:2022}, whereby the density operator is at random times reset to a given form (see later). 

We feel it pertinent at this stage to cite a number of recent work on themes related to resets of quantum dynamics. For the tight-binding Hamiltonian on an infinite line and with a detector placed on a fixed site, Refs.~\cite{ref1,ref2} studied the problem of first detected arrival of the tight-binding particle at the detector site. In particular, the authors addressed the issue of the particle, starting at a fixed site and being subject to stroboscopic measurements after every fixed interval of time $\tau$, to be detected for the first time at the detector site; Ref.~\cite{ref3} extended the study to the more general case of measurements at independent and identically-distributed random time intervals. The issue of entanglement in open and noisy quantum systems was studied in the framework of a system of interacting quantum particles, which can interact and exchange particles with the environment~\cite{ref4}. Here, the effect of decoherence is supposed to be counteracted by having the system particles randomly reset to some standard initial state, e.g., by replacing them with particles from the environment (see also Ref.~\cite{ref5}). Closed quantum many-body systems subject to stochastic resetting, whereby the system is reinitialized to a state chosen from a set of reset states conditionally on the outcome of a measurement taken immediately before resetting, were considered in Ref.~\cite{ref6}.An explicit connection was established between quantum quenches in closed systems and the emergent open system dynamics induced by stochastic resetting. 

Reference~\cite{ref7} studied the dynamics of a non-interacting spin system that is undergoing coherent Rabi oscillations in presence of stochastic resetting, showing that the latter induces long-range quantum and classical correlations in the emergent dissipative dynamics as well as in the non-equilibrium stationary state. For Markovian open quantum systems subject to stochastic resetting, the ensuing dynamics was shown in Ref.~\cite{ref8} to be non-Markovian and has the form of a generalized Lindblad equation. Interestingly, the statistics of quantum-jumps can be exactly derived. It was demonstrated that stochastic resetting may be employed as a tool towards tuning of the statistics of quantum-jump trajectories and dynamical phases of open quantum systems. In Ref.~\cite{ref9}, evolution of an arbitrary open quantum system under a resetting process was studied, and a universal behaviour for the mean return time that goes beyond unitary dynamics and Markovian measurements was unveiled. In particular, it was shown that the mean measurement time may be used as a tuning parameter to minimize the mean transition time between states of the system. Reference~\cite{ref10} studied  entanglement dynamics in monitored quantum many-body systems with well-defined quasiparticles, with the entanglement implemented by ballistically propagating non-Hermitian quasiparticles that are stochastically reset by the measurement protocol with a rate given by their finite inverse lifetime. It was shown that, depending on the spectrum of quasiparticle decay rates, different entanglement scaling and even sharp entanglement phase transitions may result.

Our work differs from the aforementioned class of work in that we study stochastic resets in a setup that models the realistic system of a nano wire, namely, the tight-binding chain. Moreover, we unveil a hitherto-unexplored physical phenomenon in this context, namely, that of a coherence-to-decoherence cross-over when the TBC dynamics has competing effects arising from the application of a time-dependent external field and from the implementation of stochastic resets at random times.

The averaged density
operator (averaged over different realizations of the dynamics detailed above) at time $t$ reads~\cite{Das:2022}
\begin{align}
&\overline{\rho}(t)= \sum_{\mu=0}^\infty \int_0^t \mathrm{d}t_\mu \int_0^{t_\mu} \mathrm{d}t_{\mu-1}\ldots \int_0^{t_3}\mathrm{d}t_2 \int_0^{t_2} \mathrm{d}t_1\nonumber \\
&\times F(t-t_\mu)\ee_+^{-\im\int_{t_\mu}^t \mathrm{d}t' \mathcal{L}(t')}\mathcal{T}p(t_\mu-t_{\mu-1})\ee_+^{-\im\int_{t_{\mu-1}}^{t_\mu} {\rm}\mathrm{d}t'\mathcal{L}(t')}\mathcal{T}\ldots
\mathcal{T}p(t_2-t_1)\ee_+^{-\im\int_{t_1}^{t_2} \mathrm{d}t'~\mathcal{L}(t')}\mathcal{T}p(t_1)\nonumber \\
&\times\ee_+^{-\im \int_0^{t_1} \mathrm{d}t'~\mathcal{L}(t')}\rho(0)\nonumber \\
&= U(t)\rho(0),
\label{eq:rho-evolution-1}
\end{align}
where $U(t)$ is a superoperator: it acts on one operator, namely, $\rho(0)$, to yield another operator given by the left hand side~\cite{Das:2022}. While further discussion on superoperator is relegated to Appendix~B, it suffices to mention the following here: The Liouville operator ${\cal L}$ associated with an ordinary operator, e.g., a time-independent Hamiltonian $H$, and when operating on an ordinary operator $A$ gives rise to the commutator of $H$ with $A$:
\begin{align}
{\cal L}\,A=[ \, H,\, A \,].
\label{eq:app1-0}
\end{align}
A Liouville operator operating on an ordinary operator gives another ordinary operator, and in this sense,  it is a superoperator. We denote the `states’ of a Liouville operator by round kets:
\begin{align}
{\cal L}|nm)={\cal L}|n\rangle \langle m|=H|n\rangle \langle m|-|n\rangle \langle m|H,
\end{align}
where the complete set of states are in our case the Wannier states, namely, the set $\{|n\rangle\}$. The projection operator $\mathcal{T}$ is also a superoperator.  In Eq.~(\ref{eq:rho-evolution-1}), the quantity $F(t)=\ee^{-\lambda t}$ is the probability
for no interaction to occur during time $t$. We remark that Eq.~(\ref{eq:rho-evolution-1}) can arrived at by utilizing the picture of quantum-jump trajectories of open quantum systems, as done in Ref.~\cite{ref8}, see Eqs. (36) and (37) and the discussions therein.

In Eq.~(\ref{eq:rho-evolution-1}), the form of the projection operator $\mathcal{T}$ should be such as to ensure that following every instantaneous reset, the density operator becomes
$\rho_+(t)=\mathcal{T}\rho_-(t)=\ket{\psi_\mathrm{r}}\bra{\psi_\mathrm{r}}$, with the trace of the operators satisfying $\mathrm{Tr}[\rho_+(t)]=\mathrm{Tr}[\rho_-(t)]=1~\forall~t$ and $\rho_-(t)$ denoting the
density operator prior to the reset. Here, $|\psi_\mathrm{r}\rangle$ is the state to which, corresponding to a reset, a complete collapse of the instantaneous state vector of the system takes place. The form of the superoperator $\mathcal{T}$ is thus the same as the superoperator $\mathcal{R}$ in Eq.~(37) of Ref.~\cite{ref8}.

On differentiating Eq.~(\ref{eq:rho-evolution-1}) with respect to $t$, we get (see Appendix~C)
\begin{align}
&\frac{\mathrm{d}\overline{\rho}(t)}{\mathrm{d}t}= -\im\mathcal{L}(t) \overline{\rho}(t) \nonumber \\
&+ F(0) \sum_{\mu=1}^\infty \int_0^{t} \mathrm{d}t_{\mu-1}\ldots \int_0^{t_3}\mathrm{d}t_2 \int_0^{t_2} \mathrm{d}t_1\nonumber \\
&\times \mathcal{T}p(t-t_{\mu-1})\ee_+^{-\im\int_{t_{\mu-1}}^{t} {\rm}\mathrm{d}t'\mathcal{L}(t')}\mathcal{T}\ldots
\mathcal{T}p(t_2-t_1)\ee_+^{-\im\int_{t_1}^{t_2} \mathrm{d}t'~\mathcal{L}(t')}\mathcal{T}p(t_1)\ee_+^{-\im \int_0^{t_1} \mathrm{d}t'~\mathcal{L}(t')}\rho(0) \nonumber \\
&+\sum_{\mu=1}^\infty \int_0^{t} \mathrm{d}t_{\mu}\ldots \int_0^{t_3}\mathrm{d}t_2 \int_0^{t_2} \mathrm{d}t_1  \, \, \left[\frac{\mathrm{d} F(t-t_\mu)}{\mathrm{d}t}\right] \ee_+^{-\im\int_{t_{\mu}}^{t} {\rm}\mathrm{d}t'\mathcal{L}(t')} \nonumber \\
&\times \mathcal{T}p(t_\mu-t_{\mu-1})\ee_+^{-\im\int_{t_{\mu-1}}^{t_\mu} {\rm}\mathrm{d}t'\mathcal{L}(t')}\mathcal{T}\ldots
\mathcal{T}p(t_2-t_1)\ee_+^{-\im\int_{t_1}^{t_2} \mathrm{d}t'~\mathcal{L}(t')}\mathcal{T}p(t_1)\ee_+^{-\im \int_0^{t_1} \mathrm{d}t'~\mathcal{L}(t')}\rho(0)  \nonumber \\
&= - \im ~\mathcal{L}(t) ~\overline{\rho}(t)  + \lambda \mathcal{T} \, \overline{\rho}(t) - \lambda \overline{\rho}(t)  \nonumber \\
&=\mathcal{W}[\overline{\rho}(t)] \, , 
\label{eq:Lindblad}
\end{align} 
where in the first equality, the first term on the right hand side denotes the contribution in absence of resets (i.e., it stands for the unitary evolution of the density operator), while the second and the third term on the right hand side stand for reset events and account for non-unitary evolution. The quantity $\overline{\rho}^{(0)}(t)$ in the second equality of Eq.~\eqref{eq:Lindblad} corresponds to the term of the density operator with no instantaneous interaction, and is defined in Eq.~\eqref{eq:rho-evolution-rho0}.
In the spirit of the Lindbladian formalism describing non-unitary evolution of the density operator in quantum mechanics, we refer to Eq.~(\ref{eq:Lindblad}) as the Lindblad master equation, and the superoperator $\mathcal{W}$ as the Lindbladian for reset dynamics~\cite{ref4,ref5,ref6,ref8,Rose}.

Considering the Laplace transform (with the Laplace transform operator denoted by the symbol $\mathfrak{L}$) of Eq.~(\ref{eq:rho-evolution-1}) and using standard results on the Laplace transform of a series of
convolutions,  one gets by following the basic steps outlined in Ref.~\cite{Das:2022} that 
\begin{align}
&\widetilde{\overline{\rho}}(s) \equiv \mathfrak{L}(\overline{\rho}(t)) = \widetilde{U}(s)\rho(0); \label{eq:rho-in-lap} \\
&\widetilde{U}(s) =\widetilde{U}_0(s)+\lambda
\widetilde{U}_0(s)\mathcal{T}\widetilde{U}_0(s)+\lambda^2 \widetilde{U}_0(s) \mathcal{T}
\widetilde{U}_0(s) \mathcal{T} \widetilde{U}_0(s)+\ldots,
\label{eq:U-expansion}
\end{align}
where we have
\begin{align}
\widetilde{U}_0(s) \equiv \int_0^\infty \mathrm{d}t~\ee_+^{-(s+\lambda)t-\im \int_0^t \mathrm{d}t'~\mathcal{L}(t')}.
\label{eq:U0s} 
\end{align}

We now specialize to the TBC subject to a periodic field,  and discuss the case of stochastic resets,  whereby the density operator is at random times reset to the form $|\mathcal{N}\rangle \langle \mathcal{N}|$ corresponding to a complete collapse of the instantaneous state vector of the system onto the state $|\mathcal{N}\rangle$ corresponding to an arbitrary site $\mathcal{N}$ on the lattice.  We take the TBC particle to be located on site $n_0$ at the start of the dynamics at time $t=0$, so that setting $\mathcal{N}=n_0$ implements reset of the density operator to its form prior to switching-on of any environmental coupling, namely, the form $|n_0\rangle \langle n_0|$. As discussed earlier, the form of the projection operator $\mathcal{T}$ should be such as to ensure that following every instantaneous reset, the density operator becomes
$\rho_+(t)=\mathcal{T}\rho_-(t)=\ket{\mathcal{N}}\bra{\mathcal{N}}$, with $\mathrm{Tr}[\rho_+(t)]=\mathrm{Tr}[\rho_-(t)]=1~\forall~t$. Following the discussion in Ref.~\cite{Das:2022}, these requirements lead to $\mathcal{T}$ of the form 
\begin{align}
(n_1 n_1'|\mathcal{T}|n_2 n_2')=\delta_{n_1,n_1'}\delta_{n_2,n_2'}\delta_{n_1,\mathcal{N}}.
\label{eq:Tmatrix}
\end{align}

To proceed, let us first obtain the matrix elements of the superoperator $\widetilde{U}_0(s)$ defined in Eq.~\eqref{eq:U0s}:
\begin{align}
(  m n |  \widetilde{U}_0(s) | m'n' ) 
&= ( mn|  \int_0^\infty \mathrm{d}t~  \ee_+^{-(s+\lambda){\mathbb{I}}t - \im \int_0^t \mathrm{d}t'~\mathcal{L}(t')} | m'n' ) \nonumber \\
&=\int_0^\infty \mathrm{d}t~\ee^{-(s+\lambda)\mathbb{I}t}\la m|\ee_+^{-\im \int_0^t \mathrm{d}t'~H(t')}|m'\ra \la n'|\ee_-^{\im \int_0^t \mathrm{d}t'~H(t')}|n\ra.
\label{eq:result1}
\end{align}
Using Eq.~(\ref{eq:HTBLt'-relation}), we get 
\begin{align}
&(  m  n |  \widetilde{U}_0(s) | m'n' ) \nonumber \\
&=\int_0^\infty \mathrm{d}t~\ee^{-(s+\lambda)\mathbb{I}t} \langle m|\ee^{-\im \int_0^t \mathrm{d}t'~H_0(t')} \ee^{-\im \int_0^t \mathrm{d}t'~H_\mathrm{TBL}(t')}|m'\rangle\langle n'|\ee^{\im \int_0^t \mathrm{d}t'~H_\mathrm{TBL}(t')}\ee^{\im \int_0^t \mathrm{d}t'~H_0(t')}|n\rangle\nonumber \\
&=\int_0^\infty \mathrm{d}t~\ee^{-(s+\lambda)\mathbb{I}t} \ee^{\im (F_0/\omega)\sin (\omega t)(n-m)} \langle m |\ee^{-\im \int_0^t \mathrm{d}t'~H_\mathrm{TBL}(t')}|m'\rangle  \langle n' |\ee^{\im \int_0^t \mathrm{d}t'~H_\mathrm{TBL}(t')}|n\rangle \nonumber \\
&=\int_0^\infty \mathrm{d}t~\ee^{-(s+\lambda)\mathbb{I}t} \ee^{\im (F_0/\omega)\sin (\omega t)(n-m)} \nonumber \\
&\times \frac{1}{(2\pi)^2}\int_{-\pi}^\pi \mathrm{d}k\int_{-\pi}^\pi \mathrm{d}k'~ \ee^{\im ((m-m')k'+(n'-n)k)} \ee^{\im \Delta \mathrm{Re}[(\ee^{\im k}-\ee^{\im k'})w(t)]}.
\label{eq:U-identity-0}
\end{align}
It then follows on using $\sum_{m = -\infty}^{\infty} \exp(\im m (k-k'))=2\pi \delta(k-k')$ that
\begin{align}
\sum_m (mm|\widetilde{U}_0(s)|nn)=\frac{1}{s+\lambda}.
\label{eq:U-identity-3}
\end{align}

Let us now calculate the probability $\overline{P}_m(t)$ for the particle to be on site $m$ at time $t$ while evolving according to our scheme of stochastic resets at random times.  Here, the overline denotes as in Eq.~(\ref{eq:rho-evolution-1}) an average over different realizations of the reset dynamics.  We have by definition that
\begin{align}
\overline{P}_m(t)=\langle m|\overline{\rho}(t)|m\rangle.
\end{align}
Denoting by $\widetilde{\overline{P}}_m(s)$ the Laplace transform of $\overline{P}_m(t)$,  Eqs.~\eqref{eq:rho-in-lap} and~\eqref{eq:U-expansion} give
\begin{align}
\widetilde{\overline{P}}_m(s) &= \la m | \widetilde{U}_0(s)\rho(0) +\lambda 
\widetilde{U}_0(s)\mathcal{T}\widetilde{U}_0(s) \rho(0)+\lambda^2 \widetilde{U}_0(s) \mathcal{T}
\widetilde{U}_0(s) \mathcal{T} \widetilde{U}_0(s) \rho(0)+\ldots |m\ra \nonumber \\
&\equiv \sum_{p = 0}^{\infty} \widetilde{\overline{P}}_m^{(p)}(s).
\label{eq:Pms}
\end{align}
We have
\begin{align}
\widetilde{\overline{P}}_m^{(0)}(s) &=\la m|\widetilde{U}_0(s)\rho(0)| m
\ra =\sum_{n,n'}(mm|\widetilde{U}_0(s)|nn')\bra{n}\rho(0)\ket{n'} = (mm|\widetilde{U}_0(s)|n_0n_0) \nonumber \\
&=\int_0^\infty \mathrm{d}t~\ee^{-(s+\lambda)t} \frac{1}{(2\pi)^2}\int_{-\pi}^\pi \mathrm{d}k\int_{-\pi}^\pi \mathrm{d}k'~ \ee^{\im ((m-n_0)(k'-k)} \ee^{\im \Delta \mathrm{Re}[(\ee^{\im k}-\ee^{\im k'})w(t)]}, 
\label{eq:P0s-1}
\end{align}
where we have used Eq.~\eqref{eq:U-identity-0} and the fact that $\rho(0)=|n_0\rangle \langle n_0|$. We thus obtain $\overline{P}_m^{(0)}(t)$ from Eq.~(\ref{eq:P0s-1}) as
\begin{align}
\overline{P}_m^{(0)}(t)&=\frac{\ee^{-\lambda t}}{(2\pi)^2}\int_{-\pi}^\pi \mathrm{d}k\int_{-\pi}^\pi \mathrm{d}k'~ \ee^{\im ((m-n_0)(k'-k)+\im \Delta \mathrm{Re}[(\ee^{\im k}-\ee^{\im k'})w(t)]} \nonumber \\
&=\frac{\ee^{-\lambda t}}{(2\pi)^2}\int_{-\pi}^\pi \mathrm{d}k\int_{-\pi}^\pi \mathrm{d}k'~ \ee^{\im (m-n_0)(k'-k)} \ee^{\im \Delta[(\cos k-\cos k')u(t)-(\sin k-\sin k')v(t)]} \nonumber \\
& = \ee^{-\lambda t}J_{m-n_0}^2(\Delta |w(t)|),
\label{eq:P0-2}
\end{align}
where we have used Eqs.~(\ref{eq:Pmt-0}) and~(\ref{eq:Pmt-final}).
 
Next, we get from Eq.~\eqref{eq:Pms} on using Eqs.~\eqref{eq:U-identity-0} and~(\ref{eq:U-identity-3}) that
\begin{align}
\widetilde{\overline{P}}_m^{(1)}(s)
&=\lambda\sum_{\substack{n_1,n_2,n_3,\\ n_4,n_5,n_6}}
(mm|\widetilde{U}_0(s)|n_1n_2)(n_1n_2|\mathcal{T}|n_3n_4)  (n_3n_4|\widetilde{U}_0(s)|n_5n_6)\la
n_5|\rho(0)|n_6\ra \nonumber \\
&=\lambda\sum_{n_3} (mm|\widetilde{U}_0(s)|\mathcal{N}\mathcal{N})(n_3n_3|\widetilde{U}_0(s)|n_0n_0) \nonumber \\
&=\lambda\int_0^\infty \mathrm{d}t~\frac{\ee^{-(s+\lambda)t}}{(s+\lambda)} \frac{1}{(2\pi)^2}\int_{-\pi}^\pi \mathrm{d}k\int_{-\pi}^\pi \mathrm{d}k'~ \ee^{\im ((m-\mathcal{N})(k'-k)} \ee^{\im \Delta \mathrm{Re}[(\ee^{\im k}-\ee^{\im k'})w(t)]} \nonumber \\
&=\lambda\int_0^\infty \mathrm{d}t~\frac{\ee^{-(s+\lambda)t}}{(s+\lambda)}J_{m-\mathcal{N}}^2(\Delta |w(t)|),
\end{align}
where we have used Eqs.~(\ref{eq:Pmt-0}) and~(\ref{eq:Pmt-final}).
Since the Laplace transform of a convolution of two functions is given by the product of their individual Laplace transforms,  we get 
\begin{align}
\widetilde{\overline{P}}_m^{(1)}(s) &=\lambda \mathfrak{L}\Big(\int_0^t \mathrm{d}t'\,\ee^{-\lambda (t-t')}\ee^{-\lambda t'}J_{m-\mathcal{N}}^2(\Delta |w(t')|)\Big)\nonumber \\
&=\lambda \mathfrak{L}\Big(\ee^{-\lambda t} \int_0^t \mathrm{d}t'\,J_{m-\mathcal{N}}^2(\Delta |w(t')|)\Big),
\end{align}
yielding
\begin{align}
\overline{P}_m^{(1)}(t)=\lambda \ee^{-\lambda t}\int_0^t \mathrm{d}t'~J^2_{m-\mathcal{N}}(\Delta|w(t')|).
\end{align} 

In a similar manner, we get
\begin{align}
\widetilde{\overline{P}}_m^{(2)}(s) &=\lambda^2\sum_{n_1,n_2}
(mm|\widetilde{U}_0(s)|\mathcal{N}\mathcal{N})(n_1n_1|\widetilde{U}_0(s)|\mathcal{N}\mathcal{N}) (n_2n_2|\widetilde{U}_0(s)|n_0n_0) \nonumber \\
&=\lambda^2\int_0^\infty \mathrm{d}t~\frac{\ee^{-(s+\lambda)t}}{(s+\lambda)^2} \frac{1}{(2\pi)^2}\int_{-\pi}^\pi \mathrm{d}k\int_{-\pi}^\pi \mathrm{d}k'~ \ee^{\im ((m-\mathcal{N})(k'-k)} \ee^{\im \Delta \mathrm{Re}[(\ee^{\im k}-\ee^{\im k'})w(t)]} \nonumber \\
&=\lambda^2\int_0^\infty \mathrm{d}t~\frac{\ee^{-(s+\lambda)t}}{(s+\lambda)^2}J_{m-\mathcal{N}}^2(\Delta |w(t)|) \nonumber \\
&=\lambda^2 \mathfrak{L}\Big(\int_0^t \mathrm{d}t'\,\ee^{-\lambda(t-t')}(t-t')~\ee^{-\lambda t'}J_{m-\mathcal{N}}^2(\Delta |w(t')|)\Big),
\end{align}
yielding 
\begin{align}
\overline{P}_m^{(2)}(t)=\lambda^2~\ee^{-\lambda t}\int_0^t \mathrm{d}t'~(t-t')J^2_{m-\mathcal{N}}(\Delta |w(t')|).
\end{align}

The general expression is thus given by
\begin{align}
\overline{P}_m^{(p)}(t)=\lambda^p ~\ee^{-\lambda t} \int_0^t \!\!{\rm
d}t'~\frac{(t-t')^{p-1}}{(p-1)!}~J^2_{m-\mathcal{N}}(\Delta
|w(t')|);~ p\in[1,\infty),
\end{align}
which on substituting in Eq.~\eqref{eq:Pms} and inverting back to the time domain gives 
\begin{align}
        \overline{P}_m(t) =\ee^{-\lambda t}J_{m-n_0}^2(\Delta |w(t)|)+ \lambda \int_0^t  {\rm
d}t'~\ee^{-\lambda t'}J_{m-\mathcal{N}}^2(\Delta |w(t')|).
\label{eq:result-rho}
\end{align}
One may check that in the absence of forcing, when one has $w(t) = t$, the above equation reduces to the result derived in Ref.~\cite{Das:2022}.

Equation~(\ref{eq:result-rho}) yields the averaged probability at time $t$ to be on site $m$ for a quantum particle starting at site $n_0$ and evolving under Hamiltonian (\ref{eq:TBC-Hamiltonian-field}) and subject to repeated stochastic resets at exponentially-distributed random times. This equation is nothing but the ``last renewal equation” for the probability $P_m(t)$ in presence of resets. Indeed, this equation could have been obtained by taking the expectation $\langle m|\rho(t)|m \rangle$ of Eq. (9) of Ref.~\cite{Mukherjee}, and by identifying the reset-free contribution in the equation as
$J^2_{m-n_0}(\Delta|w(t)|)$. It is rewarding to see that Eq.~(\ref{eq:result-rho}) could be arrived at by pursuing an independent approach, as done in this work, and this independent approach developed by us applies not just to the case of stochastic resets but also to discuss other forms of system-environment interactions, as elaborated in Ref.~\cite{Das:2022}. The mentioned last-renewal-equation approach, incidentally, is a fundamental concept in the field of stochastic resets, and has been exploited extensively in the literature~\cite{Mukherjee,ref6,ref7,ref8,ref9,ref10}.

The result~(\ref{eq:result-rho}) has the simple physical interpretation as that adduced in Ref.~\cite{Das:2022}: The first term on the right hand side arises from those realizations of evolution for time $t$ that did not involve a single reset.  The second term on the other hand arises from those realizations in which the last reset happened in the interval $[\, t',\, t'-\mathrm{d}t'\,]$, with $t'\in [\, 0, \, t \,]$. 

Summing over $m$ on both sides of Eq.~(\ref{eq:result-rho}),  and using the relations $J_{-m}(x)=(-1)^m J_m(x)$ and $1=J^2_0(x)+2\sum_{m=1}^\infty
J^2_m(x)$,  one may check that as desired, one has $\sum_{m=-\infty}^\infty \overline{P}_m(t)= \exp(-\lambda t)+ \lambda \int_0^t  \mathrm{d}t' \exp(-\lambda t')=1$.  Now, using the result $\sum_{m=-\infty}^{\infty} m J^2_{m-n_0}(x) =\sum_{m=-\infty}^{\infty} (m+n_0) \,J^2_{m}(x)=n_0\sum_{m=-\infty}^{\infty}J^2_{m}(x)= n_0$, one obtains from Eq.~\eqref{eq:result-rho} the average displacement of the particle from the initial position $n_0$ as
\begin{align} 
\overline{\mu}(t) &\equiv \sum_{m=-\infty}^{\infty} (m-n_0) ~\overline{P}_{m}(t) = (\mathcal{N} - n_0)  \lambda \int_0^t  {\rm
d}t'~\ee^{-\lambda |w(t')|} =(\mathcal{N} - n_0)(1-\ee^{-\lambda t}),
\label{eq:MEAN-lam-non0-field}
\end{align}
which for $\lambda=0$ reproduces the result~\eqref{eq:MEAN-lam0-field}.
Next,  by using the result $\sum_{m=-\infty}^{\infty} m^2 \,J^2_{m}(x) = x^2/2$, the MSD of the particle about the initial location $n_0$ is obtained from Eq.~\eqref{eq:result-rho} as
\begin{align}
\overline{S}(t) &\equiv \sum_{m=-\infty}^{\infty} (m-n_0)^2 ~\overline{P}_{m}(t)=\ee^{-\lambda t}\frac{\Delta^2 |w(t)|^2}{2}+\lambda \int_0^t \mathrm{d}t'~\ee^{-\lambda t'}\left(\frac{\Delta^2 |w(t')|^2}{2}+(N-n_0)^2\right) \nonumber \\ 
&=\ee^{-\lambda t} \frac{\Delta^2 |w(t)|^2}{2}+ \frac{\lambda \Delta^2}{2} \int_0^t \mathrm{d}t'~\ee^{-\lambda t'} \, |w(t')|^2 +  (\mathcal{N}-n_0)^2 (1-\ee^{-\lambda t}),
\label{eq:MSD-lam-non0-field}
\end{align}
which for $\lambda = 0$ reproduces correctly the result~\eqref{eq:MSD-lam0-field}. 

On taking $t \to \infty$ at a fixed $\lambda$, the MSD approaches a time-independent non-zero value
\begin{align}
\overline{S}(t\to \infty) = (\mathcal{N}-n_0)^2 +  \frac{\Delta^2}{2}\int_0^{\infty} \mathrm{d}{x} ~\ee^{-x} \, |w(x/\lambda)|^2 .
\label{eq:MSD-asymptotic}
\end{align}
We thus see a remarkable effect induced on performing stochastic resets at random times: The mean-squared displacement (MSD) of the particle about its initial location approaches at long times a time-independent value, implying thereby localization of the particle on the sites of the lattice at long times. This may be contrasted with the behaviour in the absence of resets, when the particle has an unbounded growth of the MSD in time, with no signatures of localization. This phenomenon of localization through stochastic resets has earlier been demonstrated by us in the case of the TBC without the external field~\cite{Das:2022},  and we show here that this phenomenon is more general and takes place even when the TBC is subject to a time-dependent external field. Comments are in order regarding the dependence of the result~(\ref{eq:MSD-asymptotic}) on the model parameters. Let us first note that $w(0)=0$ and that for small $t$, we have $w(t)=t$. It then follows that as $\lambda \to \infty$, we have 
\begin{align}
\overline{S}(t\to \infty) = (\mathcal{N}-n_0)^2 +  \frac{\Delta^2}{2\lambda^2}\int_0^{\infty} \mathrm{d}{x} ~\ee^{-x} \, x^2=(\mathcal{N}-n_0)^2 +  \frac{\Delta^2}{\lambda^2};~~\lambda \to \infty.
\label{eq:MSD-asymptotic-1}
\end{align}
We thus see that the larger the $\lambda$, the smaller is the long-time MSD about the initial location $n_0$. If $\mathcal{N}$ and $n_0$ coincide, whereby stochastic resets correspond to a complete collapse of the instantaneous state vector of the system onto the initial state $|n_0\rangle$, then in the limit of very frequent resets ($\lambda \to \infty$), the long-time MSD is zero. In other words, the particle is at long times completely stuck or localized at the initial state. One may draw a parallel between this behaviour and that of quantum Zeno effect, whereby a system that is subject to very frequent measurements is completely stuck in a given state~\cite{Zeno}.

Our results establish the following scenario.  In absence of stochastic resets, the TBC system subject to such an external field exhibits in general delocalization, unless one chooses the ratio of the strength to the frequency of the external field to be a zero of the zeroth order Bessel function of the first kind.  The latter is evidently difficult to realize in experiments owing to finite precision of any apparatus whatsoever that is employed to generate the external field.  This somewhat hopeless situation associated with attempts to localize the particle is remarkably circumvented by the introduction of stochastic resets, when one does not need at all to tune the strength of the external field to be of any special value but would just have to perform stochastic resets at random times.  Any amount of stochastic resets, quantified by the magnitude of the reset parameter $\lambda$, suffices to localize the particle, with the obvious caveat that smaller the value of $\lambda$, one has to wait longer for sufficient number of resets to take place that eventually localize the particle. 

\section{Numerical results}
\label{sec:numerics}

In this section, we report on numerical results for the TBC system in presence of external periodic forcing, with the system subject to stochastic resets at exponentially-distributed random times.  
In order to obtain our numerical results,  we consider a one-dimensional periodic lattice with $N$ sites.  The density operator and the Hamiltonian operator are then represented by $N \times N$ matrices, while the transition operator $\mathcal{T}$ is a matrix of dimension $N^2 \times N^2$.  A particular realization of the reset protocol over a total duration $t$ involves obtaining first the time gaps $\tau$ between successive resets, which we sample independently from the exponential distribution~(\ref{eq:ptau-exponential}) using standard techniques.  If in a given realization there are ${\mathfrak N}$ number of resets, then the $\tau$'s satisfy $\sum_{p=1}^{\mathfrak N} \tau_p < t$ and $\tau_{{\mathfrak N}+1} = t-\sum_{p=1}^{\mathfrak N} \tau_p$; for details,  see Ref.~\cite{Das:2022}.  Here, $\tau_p$ is the time interval between the $p$-th and the $(p-1)$-th reset.  The dynamical evolution, in which the particle starts initially from location $n_0$, involves the following: Starting with the initial density operator $\rho(0)$ that has only its $(n_0,n_0)$-th element nonzero and equal to unity,  we evolve $\rho(0)$ unitarily in time for time $\tau_1$ to obtain the evolved density operator as $\rho(\tau_1)=\exp(-\mathrm{i}\int_0^{\tau_1}\mathrm{d}t'~H(t'))\rho(0)\exp(\mathrm{i}\int_0^{\tau_1}\mathrm{d}t'~H(t'))$,  and then operate on it by the transition operator $\mathcal{T}$ given by Eq.~(\ref{eq:Tmatrix}). We then evolve unitarily the resulting matrix for time $\tau_2$,  then operate on it by $\mathcal{T}$, and so on.  The final step of evolution involves unitary evolution for time $\tau_{{\mathfrak N}+1}$.  Then,  the $(m,m)$-th element of the resulting matrix yields the probability for the given realization of the reset protocol for the particle to be found on site $m$ at time $t$.  We repeat the process for several realizations of the random times to obtain the averaged site-occupation probability $\overline{P}_m(t)$ for the particle to be on site $m$ at time $t$. 

\begin{figure}[!htbp]
\centering
\includegraphics[scale=1.1]{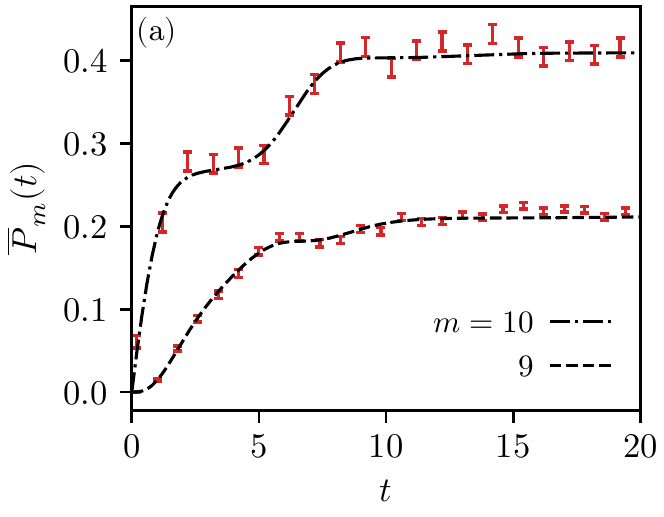}
\includegraphics[scale=1.1]{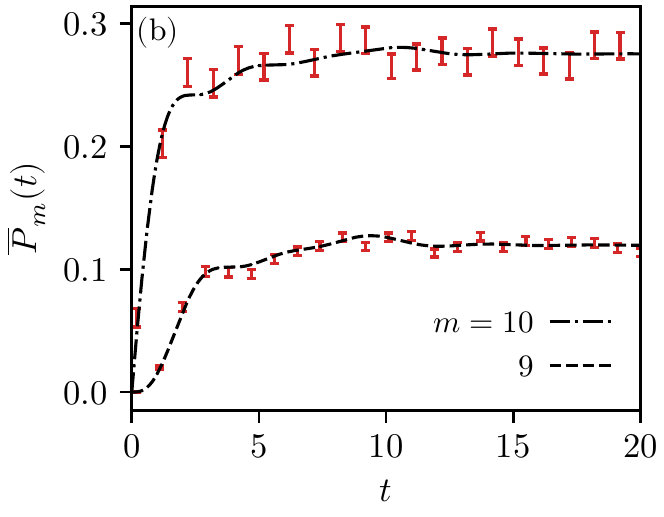}
\caption{The figures show for the field-driven TBC, Eq.~\eqref{eq:TBC-Hamiltonian-field}, subject to repeated resets at random times the dependence of the average site occupation probability $\overline{P}_m(t)$ on $t$ for two values of $m,$ namely,  $m = 9$ and $10$.  The values of the different parameters are: $\Delta= 1,  \lambda = 0.25, F_0=1.0$, while we have $\omega=0.1$ (panel (a)) and $\omega=10.0$ (panel (b)). The initial location of the particle is $n_0 = 1$, while the quantity $\mathcal{N}$ has the value $\mathcal{N}=10$ (recall that the density operator resets to the form $|\mathcal{N}\rangle\langle \mathcal{N}|$). In the plots, the line depicts the analytical result given in Eq.~\eqref{eq:result-rho}, while the points, together with error bars given by the standard deviation, are obtained from numerical implementation of the dynamics on a periodic lattice of $N = 30$ sites and involving averaging over $10^3$ realizations; see Section~\ref{sec:numerics} for details on numerical implementation. }
\label{fig:Pmt-vs-t}
\end{figure}

\begin{figure}[!htbp]
\centering
\includegraphics[scale=1.1]{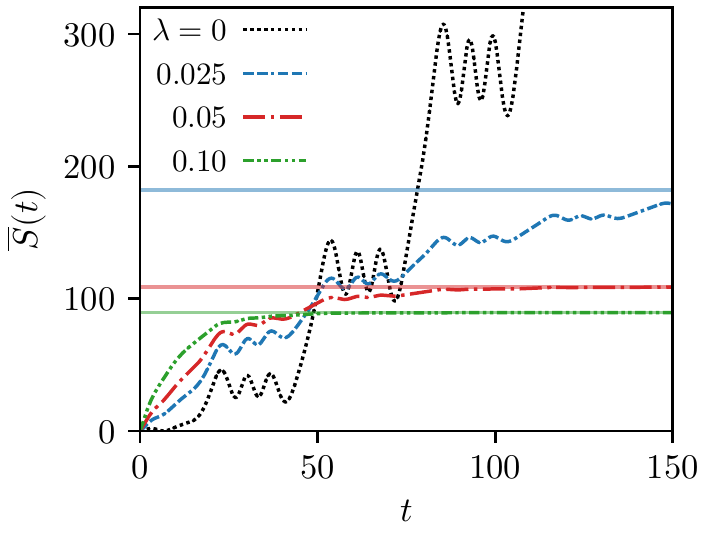}
\caption{The figure shows for the field-driven TBC, Eq.~\eqref{eq:TBC-Hamiltonian-field}, subject to repeated resets at random times the dependence of the mean-squared displacement (MSD) $\overline{S}(t)$ on time $t$.  The values of the different parameters are: $\Delta= 1,  F_0=1.0,  \omega=0.1$.  The values of the reset parameter $\lambda$ are indicated in the figure. The initial location of the particle is $n_0 = 1$, while the quantity $\mathcal{N}$ has the value $\mathcal{N}=10$ (recall that the density operator resets to the form $|\mathcal{N}\rangle\langle \mathcal{N}|$).  For each $\lambda$,  the data corresponding to the broken line are obtained by numerically evaluating Eq.~\eqref{eq:MSD-lam-non0-field} (for $\lambda=0$,  the relevant equation is Eq.~(\ref{eq:MSD-lam0-field})),  while those for the solid line represent the numerically-evaluated asymptotic value~(\ref{eq:MSD-asymptotic}).}
\label{fig:Sbar-vs-t}
\end{figure}

\begin{figure}[!htbp]
\centering
\includegraphics[scale=1.1]{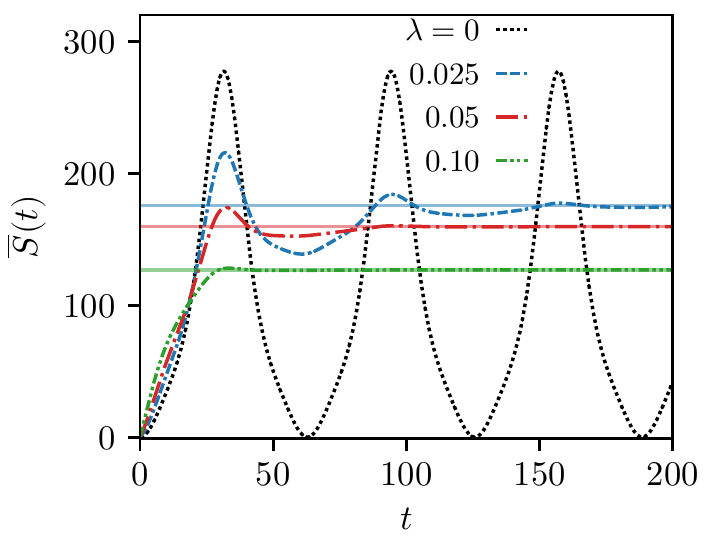}
\caption{The figure shows for the field-driven TBC, Eq.~\eqref{eq:TBC-Hamiltonian-field}, subject to repeated resets at random times the dependence of the mean-squared displacement (MSD) $\overline{S}(t)$ on time $t$.  We have here $\Delta=1$, while the values of the reset parameter $\lambda$ are indicated in the figure. The initial location of the particle is $n_0 = 1$, while the quantity $\mathcal{N}$ has the value $\mathcal{N}=10$ (recall that the density operator resets to the form $|\mathcal{N}\rangle\langle \mathcal{N}|$).  The strength of the field has the value $F_0=0.2404825558$, while its frequency has the value $\omega=0.1$, such that $F_0/\omega$ equals the first zero of $J_{0}(x)$, the zeroth order Bessel function of the first kind.  For each $\lambda$,  the data corresponding to the broken line are obtained by numerically evaluating Eq.~\eqref{eq:MSD-lam-non0-field} (for $\lambda=0$,  the relevant equation is Eq.~(\ref{eq:MSD-lam0-field})), while those for the solid line represents the numerically-evaluated asymptotic value~(\ref{eq:MSD-asymptotic}). }
\label{fig:Sbar-vs-t-Bessel0}
\end{figure}

In Fig.~\ref{fig:Pmt-vs-t}, we plot results for $\overline{P}_m(t)$ obtained by implementing the aforementioned numerical procedure by considering the number of sites $N=30$ and compare them against the analytical result~\eqref{eq:result-rho} valid in the limit $N \to \infty$.   The values of the parameters $\Delta$, $\lambda$, $F_0$, $\omega$, $N$, $n_0$, $\mathcal{N}$ are mentioned in the figure caption.  One observes, up to fluctuations induced by the use of a finite $N$, a perfect agreement between theory and numerics. In Figs.~\ref{fig:Sbar-vs-t} and~\ref{fig:Sbar-vs-t-Bessel0},  we show the dependence of the MSD $\overline{S}(t)$ on time $t$ for respectively the case in which the ratio of the strength $F_0$ to the frequency $\omega$ of the external field has a general value and when it has a value that equals the first zero of the zeroth order Bessel function of the first kind; the values of the other parameters ($\Delta,\lambda, n_0, \mathcal{N}$) are mentioned in the respective figure captions.  The data for both the figures are based on our analytical results.  Figure~\ref{fig:Sbar-vs-t} shows that for $\lambda=0$, the system shows delocalization characterized by an unbounded growth of the MSD with time, while for $\lambda \ne 0$,  the system shows clear signatures of localization induced by stochastic resets.  In Fig.~\ref{fig:Sbar-vs-t-Bessel0}, we observe dynamic localization of the bare model (namely, for $\lambda=0$,  we have $\overline{S}(t=nT)=0$) crossing over to reset-induced localization.  The foregoing observations are all in tune with our theoretical analysis and the ensuing discussions summarized in the preceding section.

\section{Conclusions}
\label{sec:conclusions}
In this work, we studied the case of the tight-binding chain (TBC) in a time-dependent external field that is subject to stochastic resets at exponentially-distributed random times.  The latter involves the density operator of the system undergoing resetting at random times to a given form, and mimics interaction of the TBC system with the external environment.  To address analytically such dynamical scenarios, we have recently developed in Ref.~\cite{Das:2022} a stochastic Liouville equation approach to study the evolution of a general density operator in the case in which its unitary evolution is interrupted at random times with interactions with the external environment.  While the explicit application of the formalism presented therein was for time-independent TBC Hamiltonian,  we show here how the formalism applies with suitable modifications also to the situation in which the Hamiltonian has an explicit dependence on time. The main theoretical result ensuing from our study of the TBC system subject to a time-dependent external field is the remarkable effect of localization of the TBC particle on the sites of the underlying lattice at long times, an effect that owes its origin solely to the protocol of stochastic resets that the system is subject to at random times. Indeed, the bare system with no reset exhibits delocalization of the particle, whereby the particle does not have a time-independent probability distribution of being found on different sites even at long times and one has a concomitant unbounded growth in time of the mean-squared displacement of the particle about its initial location.  One may induce localization in the bare model only through tuning the ratio of the strength to the frequency of the field to have a special value,  namely, equal to one of the zeros of the zeroth order Bessel function of the first kind.  While such a condition may be difficult to achieve in experiments, we showed here that localization may be induced by a far simpler procedure of subjecting the system to stochastic resets.  We feel that this effect of localization induced by stochastic resets of quantum dynamics is more general than the situation of time-independent TBC Hamiltonian explored by us in Ref.~\cite{Das:2022} and the one of time-dependent TBC system addressed in this work, and is a research direction that definitely deserves further exploration in more general scenarios, e.g., in the case of many-body interacting quantum systems. It would also be interesting to apply the formalism employed in this work to discuss the effects of stochastic resetting on the first detection probability of a quantum particle within the set-up of Refs.~\cite{ref1,ref2,ref3}. Another direction to pursue would be to apply the formalism to study Markovian open quantum systems subject to stochastic resetting~\cite{ref8}.  

\section{Acknowledgements}

SG acknowledges support from the Science and Engineering Research
Board (SERB), India under SERB-MATRICS scheme Grant No.
MTR/2019/000560, and SERB-CRG scheme Grant No. CRG/2020/000596. 
He also thanks ICTP–Abdus Salam International Centre for Theoretical Physics, Trieste, Italy, for support under its Regular Associateship scheme. SD is grateful to the Indian National Science Academy for support through their Honorary Scientist Scheme. We thank the two anonymous referees for their valuable comments that helped us improve the presentation.

\section{Appendix A: Evolution of the density operator under a time-dependent Hamiltonian}
\label{app2}

Denoting the time-dependent Hamiltonian by $H(t)$ and the instantaneous wave vector of the system as $|\psi(t)\rangle$, we have the density operator given by
\begin{align}
\rho(t)=|\psi(t)\rangle \langle \psi(t)|,
\end{align}
and the Schro\"{e}dinger time-evolution for $|\psi(t)\rangle$:
\begin{align}
\im \frac{\partial |\psi(t)\rangle}{\partial t}=H(t)|\psi(t)\rangle. 
\end{align}
The above equation may be solved iteratively as follows:
\begin{align}
\im |\psi(t)\rangle
&=\im |\psi(0)\rangle+\int_0^t \mathrm{d}t'~H(t')|\psi(t')\rangle \nonumber \\
&=\im |\psi(0)\rangle+\int_0^t \mathrm{d}t'~H(t')\left(|\psi(0)\rangle -\im \int_0^{t'}\mathrm{d}t''~H(t'')|\psi(t'')\rangle\right) \nonumber \\
&=\im |\psi(0)\rangle+\int_0^t \mathrm{d}t'~H(t')|\psi(0)\rangle-\im \int_0^t \mathrm{d}t'~H(t') \int_0^{t'} \mathrm{d}t''~H(t'')|\psi(0)\rangle+\ldots.
\end{align}
\begin{figure}[!htbp]
\centering
\includegraphics[scale=1.1]{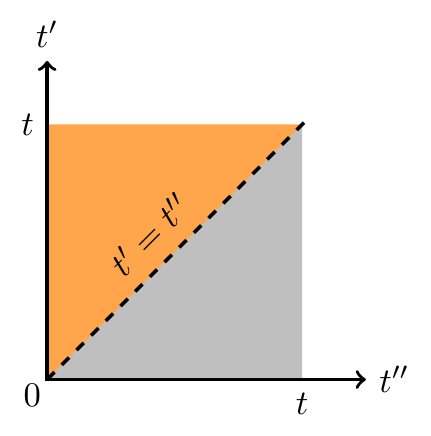}
\caption{The figure shows the orange-shaded region over which the integral $\int_0^t \mathrm{d}t' \int_0^{t'}\mathrm{d}t'' H(t')H(t'')$ and the grey-shaded region over which the integral $\int_0^t \mathrm{d}t' \int_{t'}^{t}\mathrm{d}t'' H(t')H(t'')$ are to be performed, see Appendix~A.}
\label{fig:integral}
\end{figure}
The integral $\int_0^t \mathrm{d}t' \int_0^{t'}\mathrm{d}t''~H(t')H(t'')$ has to be evaluated over the orange-shaded region shown in Fig.~\ref{fig:integral}. Writing the integral as $\int_0^t \mathrm{d}t' \int_0^{t'}\mathrm{d}t''~H(t')H(t'')=\int_0^t \mathrm{d}t' \left(\int_0^t-\int_{t'}^{t}\right)\mathrm{d}t''~H(t')H(t'')$ and noting that the area of the orange-shaded and the grey-shaded region in Fig.~\ref{fig:integral} are equal, we find that $\int_0^t \mathrm{d}t' \int_0^{t'}\mathrm{d}t''~H(t')H(t'')=(1/2)\int_0^t \int_0^t \mathrm{d}t' \mathrm{d}t''~H(t')H(t'')$ with the restriction of time ordering in view of $H(t)$ at two different times not commuting with each other: we need to first evaluate the integral over $t''$ and then the one over $t'$. The same comment applies to other integrals of the same class.  We thus obtain
\begin{align}
|\psi(t)\rangle &= 
\bigg(\mathbb{I}-\im \int_0^t \mathrm{d}t'~H(t')+\frac{\im^2}{2!}\int_0^t \int_0^t \mathrm{d}t' \mathrm{d}t''~H(t')H(t'') \nonumber \\
&\hskip20pt -\frac{\im^3}{3!}\int_0^t \int_0^t \int_0^t \mathrm{d}t' \mathrm{d}t'' \mathrm{d}t'''~H(t')H(t'')H(t''')+\ldots\bigg)|\psi(0)\rangle,
\end{align} 
and hence, that
\begin{align}
\langle \psi(t)| &= \langle \psi(0)|\bigg(\mathbb{I}+\im \int_0^t \mathrm{d}t'~H(t')+\frac{\im^2}{2!}\int_0^t \int_0^t \mathrm{d}t' \mathrm{d}t''~H(t'')H(t') \nonumber \\
&\hskip53pt +\frac{\im^3}{3!}\int_0^t \int_0^t \int_0^t \mathrm{d}t' \mathrm{d}t'' \mathrm{d}t'''~H(t''')H(t'')H(t')+\ldots\bigg).
\end{align}
We thus obtain $\rho(t)$ as
\begin{align}
\rho(t)=\ee_+^{-\im \int_0^t \mathrm{d}t'~H(t')}\,\rho(0) \, \ee_-^{\im \int_0^t \mathrm{d}t'~H(t')},
\label{eq:rho-evolution-app}
\end{align}
which is Eq.~(\ref{eq:rho-evolution}) of the main text. Here, we have invoked time ordering in writing down the exponential factors; the minus and plus subscripts on the exponential indicate negative (i.e., the latest time to the right) and positive (i.e., the latest time to the left) time ordering, respectively. Equation~(\ref{eq:rho-evolution-app}) may be used to define the Liouville-operator-evolution in the case of a time-dependent Hamiltonian as
\begin{align}
\ee_+^{-\im \int_0^t \mathrm{d}t'~\mathcal{L}(t')}\rho(0)=\ee_+^{-\im \int_0^t \mathrm{d}t'~H(t')}\,\rho(0) \, \ee_-^{\im \int_0^t \mathrm{d}t'~H(t')}.
\end{align}

 \section{Appendix B: The notion of a superoperator}
\label{app1}

We follow here Ref. \cite{Dattagupta:1987} in discussing in further detail the notion of a superoperator. As mentioned in the main text, the Liouville operator ${\cal L}$ associated with an ordinary operator, e.g., a time-independent Hamiltonian $H$, and when operating on an ordinary operator $A$ gives rise to the commutator of $H$ with $A$:
\begin{align}
{\cal L}\,A=[ \, H,\, A \,].
\label{eq:app1}
\end{align}
A Liouville operator is a superoperator that operating on an ordinary operator gives another ordinary operator. The `states’ of a Liouville operator are denoted by round kets:
\begin{align}
{\cal L}|nm)={\cal L}|n\rangle \langle m|=H|n\rangle \langle m|-|n\rangle \langle m|H,
\end{align}
where the complete set of states are in our case the Wannier states, namely, the set $\{|n\rangle\}$.                                         
The superoperator lives in a product Hilbert space, the dimension of which is the square of the space of the associated operator.  The `matrix elements' of the superoperator are labelled by four indices.  The closure property for the two-indexed states reads as $\sum_{\,m,\,n}|mn)(mn|=\mathbb{I}$.
In accordance with Eq.~(\ref{eq:app1}), we have
\begin{align}
\langle n|{\cal L}A|m\rangle=\sum_{m',n'}(nm|{\cal L}|n'm')\langle n'|A|m'\rangle,
\label{eq:superoperator}
\end{align}
with
\begin{align}
(nm|{\cal L}|n'm')=\langle n |H|n'\rangle \delta_{mm'}-\langle m'|H|m\rangle \delta_{nn'}.
\label{eq:matrix-elements-app}
\end{align}
The Heisenberg time-evolution of an operator may be defined as
\begin{align}
A(t)=\ee^{{\rm i}{\cal L}t}A(0)=\ee^{{\rm i}Ht}A(0)\ee^{-{\rm i}Ht},
\end{align} 
while that of the density operator is given by
\begin{align}
\rho(t)=\ee^{-{\rm i}{\cal L}t}\rho(0)=\ee^{-\im Ht}\rho(0)\ee^{\im Ht}.
\label{eq:liouville-eqn-app}
\end{align}
The rule given by Eq.~\eqref{eq:superoperator} applies to any superoperator,  and applying it to the density operator $\rho(t)$ in Eq.~(\ref{eq:liouville-eqn-app}) gives
\begin{align}
\langle m |\rho(t)|n\rangle&=\sum_{m', \, n'}(mn|\ee^{-{\rm i}{\cal L}t}|m'n')\langle m'|\rho(0)|n'\rangle 
=\sum_{m',\, n'}\langle m|\ee^{-{\rm i}Ht}|m'\rangle\langle m'|\rho(0)|n'\rangle \langle n'|\ee^{{\rm i}Ht}|n\rangle.
\label{eq:rho-matrix-element}
\end{align}
On comparing the first and the second equality in Eq.~(\ref{eq:rho-matrix-element}), we get the matrix elements of 
the superoperator $\exp({-\im \mathcal{L}t})$ as
\begin{align}
( mn| \ee^{-\im \mathcal{L}t} | m'n' ) = \la m | \ee^{-\im H t} |m'\ra \la n' | \ee^{\im H t} |n\ra . \label{eq:iLt-elements}
\end{align}
For the case of a time-dependent Hamiltonian considered in Appendix~A, we have 
\begin{align}
( mn| \ee_+^{-\im \int_0^t \mathrm{d}t'~\mathcal{L}(t')} | m'n' ) = \la m | \ee_+^{-\im \int_0^t \mathrm{d}t'~H(t')} |m'\ra \la n' | \ee_-^{\im \int_0^t \mathrm{d}t'~H(t')} |n\ra, \label{eq:iLt-elements-1}
\end{align}
which has been used in obtaining Eq.~(\ref{eq:result1}) of the main text.

\section{Appendix C: The Lindblad master equation}
\label{app3}

The averaged density operator at time $t$ may be written from Eq.~\eqref{eq:rho-evolution-1} as
\begin{align}
\overline{\rho}(t)= \sum_{\mu=0}^{\infty} \overline{\rho}^{(\mu)}(t) \, , \label{eq:rho-evolution-rho-sum}
\end{align}
where the quantity $\overline{\rho}^{(\mu)}(t)$ corresponds to the term associated with $\mu$ number of instantaneous interactions.
We then have
\begin{align}
\overline{\rho}^{(0)}(t) &\equiv ~F(t)\ee_+^{-\im \int_0^{t} \mathrm{d}t'~\mathcal{L}(t')}\rho(0) \, , \label{eq:rho-evolution-rho0} \\
\overline{\rho}^{(1)}(t) &\equiv \int_{0}^{t} \dd{t_1} F(t-t_1) ~\ee_+^{-\im \int_{t_1}^{t} \mathrm{d}t'~\mathcal{L}(t')} G_1(t_1) \, , \label{eq:rho-evolution-rho1}\\
\overline{\rho}^{(2)}(t) &\equiv \int_{0}^{t} \dd{t_2} F(t-t_2) ~\ee_+^{-\im \int_{t_2}^{t} \mathrm{d}t'~\mathcal{L}(t')} G_2(t_2) \, , \label{eq:rho-evolution-rho2}\\
&\vdots \, , \nonumber 
\end{align}
with
\begin{align}
G_1(t) &\equiv \mathcal{T} p(t) ~\ee_+^{-\im \int_0^{t} \mathrm{d}t'~\mathcal{L}(t')} \rho(0) \, , \label{eq:rho-evolution-g1}\\
G_2(t) &\equiv  \mathcal{T} \int_{0}^{t} \dd{t_1}  p(t-t_1) ~\ee_+^{-\im \int_{t_1}^{t} \mathrm{d}t'~\mathcal{L}(t')} G_1(t_1) \, , \label{eq:rho-evolution-g2}\\
&\vdots \, . \nonumber
\end{align}
Taking the derivative with respect to $t$, one obtains from Eqs.~\eqref{eq:rho-evolution-rho0}--\eqref{eq:rho-evolution-rho2} that
\begin{align}
\dv{t}( \overline{\rho}^{(0)}(t) ) &= - \im ~\mathcal{L}(t) ~\overline{\rho}^{(0)}(t)+\frac{\mathrm{d}F(t)}{\mathrm{d}t}\ee_+^{-\im \int_0^t \mathrm{d}t'~\mathcal{L}(t')}\rho(0) \, ,  \label{eq:rho-evolution-dv0}\\
\dv{t}( \overline{\rho}^{(1)}(t) ) &= - \im ~\mathcal{L}(t) ~\overline{\rho}^{(1)}(t) + F(0) G_1(t) + \int_{0}^{t} \dd{t_1} \dv{F(t-t_1)}{t}  ~\ee_+^{-\im \int_{t_1}^{t} \mathrm{d}t'~\mathcal{L}(t')} G_1(t_1) \, , \label{eq:rho-evolution-dv1}\\
\dv{t}( \overline{\rho}^{(2)}(t) ) &= - \im ~\mathcal{L}(t) ~\overline{\rho}^{(2)}(t) + F(0) G_2(t) + \int_{0}^{t} \dd{t_2} \dv{F(t-t_2)}{t}  ~\ee_+^{-\im \int_{t_2}^{t} \mathrm{d}t'~\mathcal{L}(t')} G_2(t_2) \, . \label{eq:rho-evolution-dv2}
\end{align}
Using the above equations together with Eqs.~\eqref{eq:rho-evolution-rho-sum},~\eqref{eq:rho-evolution-g1}--\eqref{eq:rho-evolution-g2},  and considering the method of induction, one obtains the first equality of Eq.~\eqref{eq:Lindblad} in the main text.
In our case with $p(t)=\lambda \exp(-\lambda t)$ and $F(t) = \exp(-\lambda t)$, one has $F(0)=1$, 
$ p(t-t_\mu)= \lambda F(t-t_\mu) = - \dd{F(t-t_\mu)}/\dd{t}  $,
and
\begin{align}
G_1(t) &= \lambda \mathcal{T} \overline{\rho}^{(0)}(t) \, , \\
G_2(t) &= \lambda \mathcal{T} ~\overline{\rho}^{(1)}(t) \, , \\
&\vdots \, . \nonumber
\end{align}
Then, Eqs.~\eqref{eq:rho-evolution-dv0}--\eqref{eq:rho-evolution-dv2} yield
\begin{align}
\dv{\overline{\rho}(t)}{t} &= - \im ~\mathcal{L}(t) ~\overline{\rho}(t) + \lambda \mathcal{T} \overline{\rho}(t)-\lambda \overline{\rho}(t) \, .
\end{align}
which is the second last equality of Eq.~\eqref{eq:Lindblad} in the main text.


\end{document}